\documentstyle[floats,aps]{revtex}
\begin{document}
\draft
\catcode`\@=11
\catcode`\@=12
\twocolumn[\hsize\textwidth\columnwidth\hsize\csname%
@twocolumnfalse\endcsname
\title{Mutual Exclusion Statistics Between Quasiparticles in the
Fractional Quantum Hall Effect}
\author{Wu-Pei Su$^1$, Yong-Shi Wu$^2$ and Jian Yang$^1$}
\address{$^1$~Department of Physics and Texas 
Center for Superconductivity,
University of Houston, Houston, Texas 77204\\
$^2$~Department of Physics, University of Utah, 
Salt Lake City, Utah 84112}

\date{Received:}
\maketitle
\begin{abstract}
In this paper we propose a new assignment for mutual 
exclusion statistics between quasielectrons and quasiholes 
in the fractional quantum Hall effect. In addition to
providing numerical evidence for this assignment, we 
show that the physical origin of this mutual 
statistics is a novel hard-core constraint due to correlation
between the distinguishable vortex-like quasiparticles.  

\end{abstract}

]

A two-dimensional electron system in a strong 
magnetic field exhibits the fractional quantum 
Hall effect (FQHE)\cite{Tsui,FQHE} at certain 
``magic'' fillings, say $\nu=1/m$ ($m$ an odd integer). 
One fascinating aspect of this effect 
is that quasiparticles are predicted
to have exotic quantum numbers. For example, 
a quasihole (QH) or a quasielectron (QE) in 
these systems (called the Laughlin $1/m$
fluids) has only a fraction, $\pm 1/m$, of 
the electron charge \cite{Laughlin}. If two 
QH's (or QE's) are exchanged, their wave function 
acquires a fractional phase $\pm \pi /m$ 
\cite{Halperin,Arovas}. 

Recently it has further been argued 
that these quasiparticles exhibit 
exotic quantum statistical\cite{Hald1} and 
thermodynamic\cite{Wu} behavior, due to a novel
rule for state counting: The total 
number of states with $N_{-}$ QH's and 
$N_{+}$ QE's is given by 
\begin{eqnarray} 
W = 
\left(\begin{array}{c}
D_+ + N_+ -1\\
N_+
\end{array}\right) 
\left(\begin{array}{c}
D_- + N_- -1\\
N_-
\end{array}\right), 
\label{counting}
\end{eqnarray}
together with ($ i,j = -$ for QH, 
$+$ for QE)
\begin{equation} 
D_i = \frac{N_{\phi}}{m} 
- g_{ii} (N_{i}-1) 
- \sum_{j\neq i} \, g_{ij} N_j\; .
\label{rule}
\end{equation}
Here $N_{\phi}$ is the external magnetic flux in 
units of $h/e$, and $N_{\phi}/m$ is the Landau 
degeneracy for the quasiparticles due to their 
fractional charge. Though the formula (\ref{counting}) 
looks formally the same as that for bosons, 
the rule (\ref{rule}) is unusual, because
the number, $D_{i}$, of available single-particle 
states for species $i$ can be, as first suggested 
by Haldane\cite{Hald1}, linearly dependent 
on the particle numbers. The coefficients 
$g_{ij}$ describe statistical exclusion between 
particles in occupying single-particle
states and, therefore, are called exclusion statistics. 
If $g_{ii}= 0$ or $1$ and $g_{ij}=0$ ($i\neq j$), 
the particle is a boson or fermion. 
The diagonal statistics $g_{ii}$ in the 
Laughlin $1/m$ fluid has previously been
determined\cite{Hald1,HXZ,jian,YangSu,CJ} to be
\begin{equation}
g_{--}= 1/m, \;\;\;\;  g_{++}=2-1/m\; .
\label{diagonal}
\end{equation}
That $g_{++}$ is $2-1/m$ rather than $-1/m$
is due to a hard-core constraint between QE's
\cite{HXZ,jian}. A distinct feature of the new rule 
(\ref{rule}) is that it naturally allows for 
the possibility of {\it mutual statistical 
exclusion} ($g_{ij}\neq 0$) between 
{\it different} species $(i\neq j)$. 
In this Letter we propose, 
and present numerical evidence to
support, that indeed this happens to the FQHE 
quasiparticles with off-diagonal 
\begin{equation}
g_{-+}=- (2-1/m), \;\;\;\; g_{+-} = 2-1/m\; ,
\label{non-diag}
\end{equation}
and show that this is due to a hard-core 
constraint between QE and QH\@. Note 
the mutual exclusion statistics in FQHE are 
{\it anti-symmetric}, in contrast to mutual exchange 
statistics which is always symmetric\cite{comm}.  

Let us first recall the arguments for the 
fractional diagonal statistics (\ref{diagonal}).
For a Laughlin liquid, the electron 
number $N_e$ and quasiparticle numbers 
$N_-$ and $N_+$ satisfy
\begin{equation}
N_{\phi} = m(N_e-1)+ N_- - N_+.
\label{flux}
\end{equation}
If the quasiparticles, viewed as vortices\cite{HaldWu}
in the Laughlin liquid, are uncorrelated,
the Hilbert-space dimension for a single vortex 
is determined by the number of fluid 
particles : $D_{\pm}=N_e+1$. Eliminating $N_e$ with 
the help of eq. (\ref{flux}), one obtains an 
expression\cite{comm1} of $D_{\mp}$ of the form
of eq. (\ref{rule}), which gives the statistics 
parameters\cite{Hald1}
\begin{equation}
g^{0}_{--} = - g^{0}_{++} 
= g^{0}_{+-} =-g^{0}_{-+}
= 1/m\; .
\label{naive}
\end{equation}
This argument needs to be refined,
if the vortices are correlated to each other; 
in such cases, one has instead
\begin{eqnarray}
D_+ = N_e+1 - \alpha_{++} (N_+ -1) 
- \alpha_{+-} N_- \; ,\nonumber \\
D_- = N_e+1 - \alpha_{--} (N_- -1) 
- \alpha_{-+} N_+\; ,
\label{alphadef}
\end{eqnarray}
with a linear dependence of $D_{i}$ on the 
excitation numbers $N_{j}$. This will
modify the statistics matrix to
\begin{equation}
g_{ij} = g_{ij}^0 + \alpha_{ij}.
\label{modify}
\end{equation}
Previously, it has been shown\cite{HXZ,jian,YangSu,CJ} that 
due to a hard-core constraint for QE's 
(not for QH's), one should assign
\begin{equation}
\alpha_{++} = 2\, , \;\;\; \alpha_{--}=0 \; .
\label{diagalpha}
\end{equation}
Thus, $g_{++}$ should be modified to that 
given by eq. (\ref{diagonal}). 

Recently we have numerically 
determined the off-diagonal $\alpha_{\pm\mp}$ 
for small systems
of interacting electrons on a sphere, 
with a magnetic monopole at its center\cite{Hald2} 
providing a total flux $N_\phi=2S$. 

The simplest case with coexisting QE and QH
is the magnetic roton band with $N_- =N_+ =1$, 
just above the $\nu=1/m$ ground state. In Fig.1, we 
present the energy spectrum of $N_e=6$ electrons 
with $2S=15$, which according to eq. (\ref{flux})
corresponds to $m=3$ and $N_- =N_+$. The unique 
ground state (with $N_-=N_+=0$) is seen 
well-separated by a gap from the 
excited states. The low-lying excited states 
above it form a visible second band and is thought 
of as containing a pair of QE and QH. To identify 
states of such quasiparticle composition \cite{jian1},
one first consider the subspace spanned 
by wave functions of the form $S^+(\alpha_0,\beta_0) 
S^-(\alpha_1,\beta_1) \Psi_m$,
where
\begin{equation}
\Psi_{m} = \prod_{i<j}(u_iv_j-v_iu_j)^{m}            
\end{equation}
is the Laughlin wave function on the sphere,
$(u,v)$ and $(\alpha,\beta)$ are the spinor 
variables describing electron and quasiparticle 
coordinates, and the operators 
\begin{eqnarray}
S^+(\alpha,\beta) = &
\prod_{j=1}^N & (\beta^* \frac{\partial}{\partial u_j}
-\alpha^* \frac{\partial}{\partial v_j})\;,\nonumber\\
S^-(\alpha,\beta) = &
\prod_{j=1}^N & (\beta u_j -\alpha v_j), 
\label{operator}
\end{eqnarray}
are respectively the QE and QH creation operators. 
By diagonalizing the Hamiltonian in this subspace
and by inspecting both angular momentum and energy, 
one is tempted to identify the states of a 
QE-QH pair to be those with a long bar in the exact 
spectrum Fig. 1. Notice that the 
$L=0$ state in this subspace actually is 
the ground state, so it should not be counted as a
true QE-QH pair state. Moreover the lowest states 
with $L=1$ obtained this way are clearly in 
the continuum above the well-separated second 
band, suggesting they do not really belong to the 
true subspace of one QE-QH pair. Thus, one should
exclude these four states (with $L=0,1$) from 
the QE-QH pair subspace; the remaining states 
stay between two dotted lines in Fig. 1. 
If we change $N_e$, while keeping $\nu=1/3$ fixed, 
numerical data always show the missing of 
four states in the magnetic roton band. 
This is an indication of mutual exclusion 
between QE and QH. Indeed, the number 
of states in the QE-QH subspace is 
$(N_e+1-\alpha_{+-})(N_e+1-\alpha_{-+})$.
With $\alpha_{+-} = \alpha_{-+} = 0$, there 
should be $(N_{e}+1)^2$ states in the second 
band. Four states missing implies 
\begin{equation}
\alpha_{+-}= -\alpha_{-+} = \pm 2\, .
\label{nondalpha}
\end{equation}

To resolve the sign ambiguity, we  
study larger systems. We choose $N_e=6$ 
as before, but add one or two extra flux quanta
(i.e.\  2S=16 or 17). Hereafter the two systems 
will be referred to as $\frac{1}{3}+$ and 
$\frac{1}{3}++$ respectively.  Their energy spectra 
are shown in Figs. 2(a) and 2(b). According to eq. 
(\ref{flux}), the systems have respectively $N_-=N_+ +1$ 
and $N_-=N_+ +2$, so the minimum number of QH's 
is 1 and 2 respectively. In Fig. 2(a), the lowest 
energy band consists of a single multiple with $L=3$, 
correspond to states with a single QH; in Fig. 2(b), 
four $L$-multiplets of lowest energies form the lowest
band, correspond to states with $2$ QH's. The state
counting for these bands agrees with the statistics 
$\alpha{--} = 0$ or $g_{--} = 1/3$. In order to study 
mutual statistics, one needs to examine higher-energy 
states which contain one more QE and QH. To properly 
identify these states, we invoke microscopic
Laughlin wave functions with appropriate quasiparticle 
composition, i.e.\ to consider the 
subspace spanned by $S^+(\alpha_0,\beta_0) 
S^-(\alpha_1,\beta_1) S^-(\alpha_2,\beta_2) \Psi_m $ 
for the $\frac{1}{3}+$ system and $S^+(\alpha_0,\beta_0) 
S^-(\alpha_1,\beta_1) S^-(\alpha_2,\beta_2) 
S^-(\alpha_3,\beta_3)\Psi_m$ for the $\frac{1}{3}++$ 
system. We first project the quasiparticle-coordinate
dependence down to the ``lowest Landau level (LLL)'' 
(with electrons as sources of quantized flux for the 
quasiparticles ), resulting in a basis of many-electron 
wave functions in this subspace\cite{comm2}. Then we 
diagonalize the Hamiltonian in this basis for the two 
systems respectively, and calculate the overlaps 
of states thus obtained with the corresponding exact 
states, whose energies are marked again by long 
bars in Fig. 2. 

Let us first examine the $\frac{1}{3}+$ system more
carefully. The above construction amounts to
having $\alpha_{+-}=\alpha_{+-}=0$, and gives $196$
states in $20$ multiplets, corresponding to those 
with a long bar in Fig. 2(a). In Table I, we see 
that $16$ multiplets of them have fairly large 
overlaps with the exact eigenstates, while $4$ 
multiplets (with $L=1,2,3,4$) with  
higher energies have small overlaps. The latter
four multiplets, together with the lowest-energy 
states at $L=3$ that actually corresponds to a single 
QH, are expected not to belong to the true 
$(N_-=2, N_+=1)$ subspace. In total $31$ states 
should be excluded: We are left with only $165$ 
states, exactly what is predicted by the formulas 
(\ref{counting}) and (\ref{alphadef}), 
with $\alpha_{--} = 0$ and
\begin{equation}
\alpha_{+-} = -2 \; , \;\;\;\; \alpha_{-+} = 2 \; .
\label{correct}
\end{equation}

For the $\frac{1}{3}++$ system, based on the 
same procedure and reasoning as in last two
paragraphs, we have identified a total of $133$ 
exact states, marked by a long bar outside the 
two dashed lines in Fig. 2(b), which all have 
small overlaps with the Laughlin quasiparticle 
wave functions corresponding to $N_-=3$ and $N_+=1$. They 
should not belong to the true ($N_-=3, N_+=1$) 
subspace, while those with a long bar 
between the two dashed 
lines have large wave function overlaps and thus 
do belong to it. The total number of states
of the latter is $588 - 133 = 455$, again exactly 
what is predicted by the formulas (\ref{counting}) 
and (\ref{alphadef}), with $\alpha_{--} = 0$ and 
$\alpha_{+-} = - \alpha_{-+} = -2$.

Thus, as far as state counting is concerned, 
the off-diagonal parameters (\ref{correct}) 
are verified by our numerical data. From eqs. 
(\ref{naive}) and (\ref{modify}), we have derived
the mutual statistics parameters (\ref{non-diag}).

To understand the origin of this statistical exclusion
between QE and QH,  
we need a highly non-trivial mechanism as it
increases the Hilbert-space dimension for a single
QE as more QH's are added, while decreases the 
Hilbert-space dimension for a single QH as more QE's 
are added. It is amusing to see that a simple 
hard-core constraint between QE and QH can achieve
just that. 

Recall that in the subspace having one QE and  
several QH's, the wave functions that describe 
statistically independent QE and QH's are those 
obtained by applying $S^+(\alpha_0,\beta_0)$ and 
$S^-(\alpha_i,\beta_i)$ to the Laughlin ground 
state $\Psi_m$. To construct a many-electron 
basis (not necessarily orthogonal) in this 
subspace, one may integrate over quasiparticle 
coordiantes as follows \cite{YangSu}: 
\begin{eqnarray}
\int d\Omega(\alpha_0,\beta_0) 
\phi_{2S_{+},k_{0}}^+(\alpha_0,\beta_0) 
S^+(\alpha_0,\beta_0) \nonumber \\
\int {\prod}_q d\Omega(\alpha_q,\beta_q) 
\phi_{2S_{-},k_{q}}^-(\alpha_q,\beta_q)              
S^-(\alpha_q,\beta_q) \Psi_m
\label{integral}
\end{eqnarray}
where $(\alpha,\beta)=(cos(\theta/2)e^{i\phi/2},
\, sin(\theta/2)e^{-i\phi/2})$, and
$d\Omega(\alpha,\beta)=sin\theta d\theta d\phi$; 
$\phi_{2S_{+}, k}^+(\alpha,\beta)$ and  
$\phi_{2S_{-},k}^-(\alpha,\beta)$ are the 
single-quasiparticle wave function in the 
LLL with total flux $2S_{+}$ and $-2S_{-}$ 
respectively, with $k$ an integer between
$1$ and $2S_{\pm}+1$ labeling
the LLL states. One obtains explicitly
\begin{equation}
\delta_{2S_+,N_{e}} \delta_{2S_-, N_{e}}\; 
g_{k_0}^+ {\prod}_q  g_{k_q}^- \Psi_m
\end{equation}
where
\begin{eqnarray}
g_{k}^{+}=(-1)^{k-1}
[N_{e}!/(N_{e}-k+1)!(k-1)!]^{-\frac{1}{2}}  \nonumber \\
\sum_{1\leq l_1<l_2 \cdot\cdot\cdot 
<l_{N_{e}-k+1}\leq N_{e}}
\frac{\partial}{\partial v_{l_1}}
\frac{\partial}{\partial v_{l_2}}\cdot\cdot\cdot
\frac{\partial}{\partial v_{l_{N_{e}-k+1}}}
\prod_{l(\neq l_i)} \frac{\partial}{\partial u_l}
\end{eqnarray}
\begin{eqnarray}
g_{k}^{-}=
[N_{e}!/(N_{e}-k+1)!(k-1)!]^{-\frac{1}{2}}    \nonumber \\
\sum_{1\leq l_1<l_2 \cdot\cdot\cdot <l_{k-1}\leq N_{e}}
v_{l_1}v_{l_2}
\cdot\cdot\cdot v_{l_{k-1}}
\prod_{l(\neq l_i)} u_l
\end{eqnarray}
The number of these basis functions gives 
the dimension of the subspace in agreement with eq. (6).

In the integral (\ref{integral}),
a QE can be on top of a QH. 
This can be avoided, in the spirit of
ref. \cite{HXZ}, by inserting a hard-core 
Jastrow factor between QE and QH \cite{jian}
\begin{equation}
\prod_q (\alpha_0^* \beta_q^* - \alpha_q^* \beta_0^*)^l\;, 
\label{Jastrow}
\end{equation}
with $l$ being a positive integer, into the integrand.
Expanding this Jastrow factor,
one can carry out the integration over quasiparticle coordinates
and obtain the following basis functions
\begin{eqnarray}
\delta_{2S_+, N_{e}+lN_-} 
\delta_{2S_-, N_{e}-l}
(-1)^{\sum k_q} \frac{\prod C^-_{2S_-,k_q}}{C^+_{2S_+,k_0}}
\sum_{m_1,m_2, \cdot\cdot\cdot, m_n} \nonumber \\
\prod_{q} \left( \begin{array}{c} l\\m_q \end{array} \right) 
(-1)^{\sum m_q} g^+_{k_0-\sum m_q}
\prod_q \frac{g^-_{k_q+m_q}}{(C^+_{N_{e},k_q+m_q})^2}
\Psi_m
\label{wvfnctn}
\end{eqnarray}
where $C^+_{N,k} = (-1)^{N-k+1} C^-_{N,k} =
[\frac{N+1}{4 \pi} \frac{N!}{(N-k+1)!(k-1)!}]^{\frac{1}{2}}$.
These new basis functions, 
which are linear superpositions of the old basis functions
$g^+_{k_0}\prod_q g^-_{k_q}\Psi_m$, 
are non-vanishing only when
\begin{eqnarray}
2S_+ = N_e + lN_-\, , \;\;\; 
2S_- = N_e-l\; . 
\end{eqnarray}
Here $D_{\pm}=2S_{\pm}+1$ gives the degeneracy for 
species $i$. If we set $l=2$, the dimension of 
the subspace spanned by these basis functions 
is precisely that required by 
$\alpha_{+-} = - \alpha_{-+} = -2$.
This shows that the 
physics behind the mutual statistics (\ref{non-diag}) 
is the hard-core constraint between QE and QH, 
i.e.\  the insertion of 
$\prod_q (\alpha_0^* \beta_q^* - \alpha_q^* \beta_0^*)^2$.
(The conjugate $\prod_q (\alpha_0 \beta_q 
- \alpha_q \beta_0)^2$ would not work.) 

The Jastrow factor inserted amounts effectively to attaching the same
number of flux quanta with the same sign to each quasiparticle (either
QE or QH). Since the total flux seen by a QE and a QH before imposing
the constraint are the same in magnitude and opposite in sign, the net
flux seen by the QE and the QH after imposing the constraint is
increased and decreased respectively, which results in the increase
(decrease) of the QE (QH) Hilbert-space dimension.

We still need to check that the hard-core modified 
wave functions indeed provide a good description 
for the exact states we have identified. To this end, 
we have diagonalized the Hamiltonian in the subspace 
spanned by basis functions (\ref{wvfnctn}) with $l=2$,  
as well as by basis functions with only QH's present. 
(The latter is included to get better overlaps. 
The detail will be published elsewhere.)
By matching the quantum numbers and energies of
thus-obtained states with exact states, we found 
that besides the states in the lowest band 
corresponding to purely QH states, they 
correspond to the exact states between the two dashed 
lines in Fig. 2. Moreover, their overlaps
with these exact states are all fairly large,
as can be seen from Table II. These strongly support 
the correctness of our hard-core constraint and, 
as a result, that of the mutual statistics given
in eq. (\ref{correct}) or eq. (\ref{non-diag}).
Implications of the mutual statistics to thermodynamic 
properties of FQHE quasiparticles will be published
\cite{HKWY}.

We would like to thank C.S. Ting and F.C. Zhang 
for discussions. Work at Houston was partially 
supported by the Texas 
Advanced Research Program under grant number 003652-183 
and by the Texas Center for Superconductivity at 
the University of Houston. WPS also acknowledges partial 
support from Robert A. Welch Foundation and Dornors of the 
Petroleum Research Fund administered by the 
American Chemical Society. YSW is partially supported 
by grant NSF PHY-9308543.

\begin{center}
{\bf Figure captions}\\
\end{center}
Fig.1 Energy spectrum of six electrons at $\nu=1/3$. 
The long bars represent the exact states whose 
angualr momentum and energy are compatible 
with states in the uncorrelated QE-QH scheme. 
However, the states which are beyond the dotted 
lines actually do not belong to the real QE-QH 
subspace.

Fig.2 Same as Fig. 1 for (a) the $\frac{1}{3}+$ and (b)
the $\frac{1}{3}++$ system.

\begin{center}
{\bf Table caption}\\
\end{center}
Table I: Overlaps between the exact states, with 
energy increasing from top to bottom, and 
the corresponding states in the uncorrelated QE-QH
scheme for (a) the $\frac{1}{3}+$  and (b)
the $\frac{1}{3}++$ system.
 
Table II: Same as Table I, but the uncorrelated QE-QH
scheme is replaced by the hard-core modified scheme.



\begin{references}
\bibitem{comm1} From eq. (5) on, all formulas for
state counting are in spherical geometry, for which  
eq. (2) contains constants of order unity, irrelevant 
in the thermodynamic limit.
\bibitem{comm2} 
Some details of this procedure will be shown below, 
when we discuss how to impose the hard-core constraint.
\end{references}
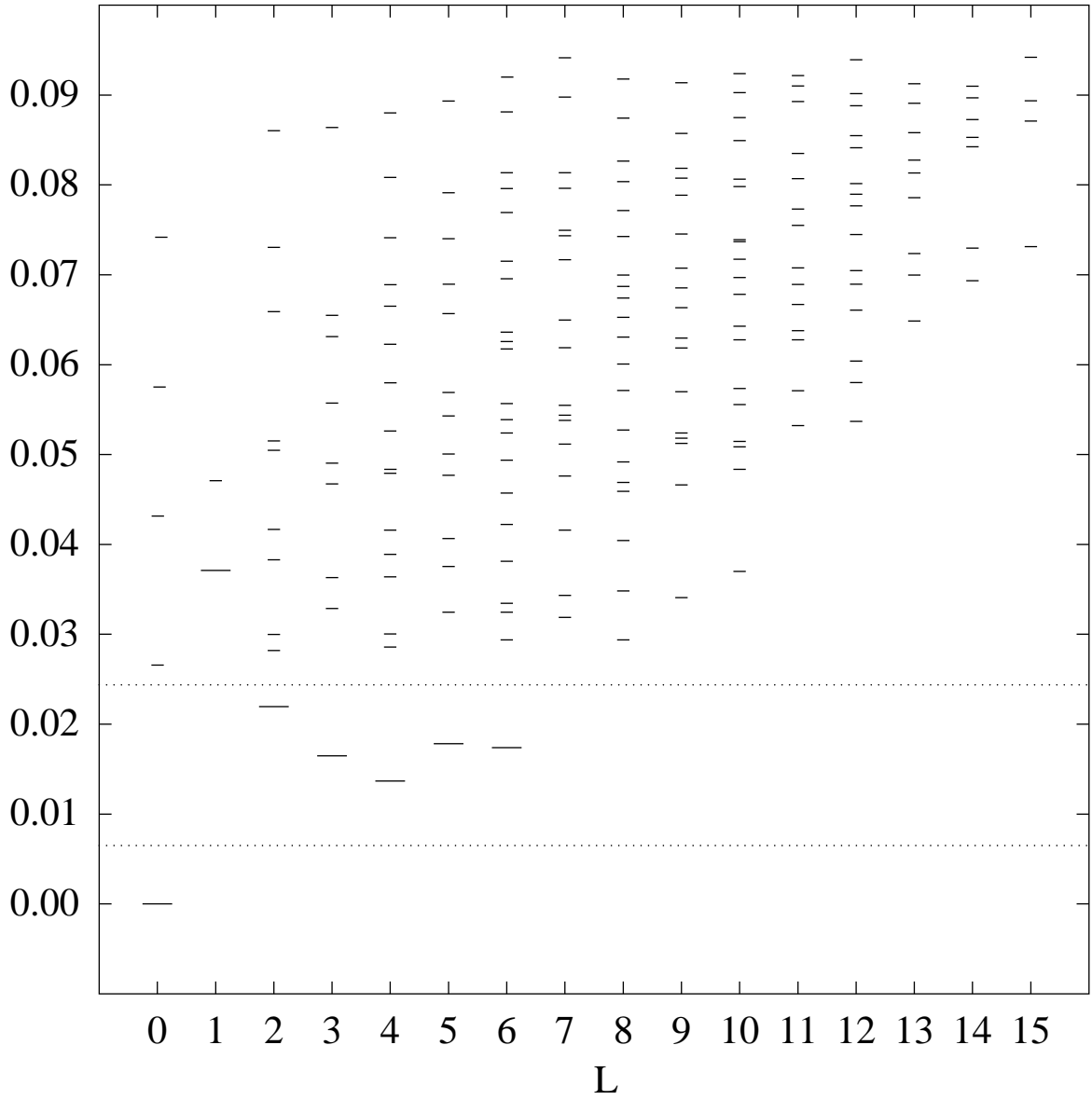
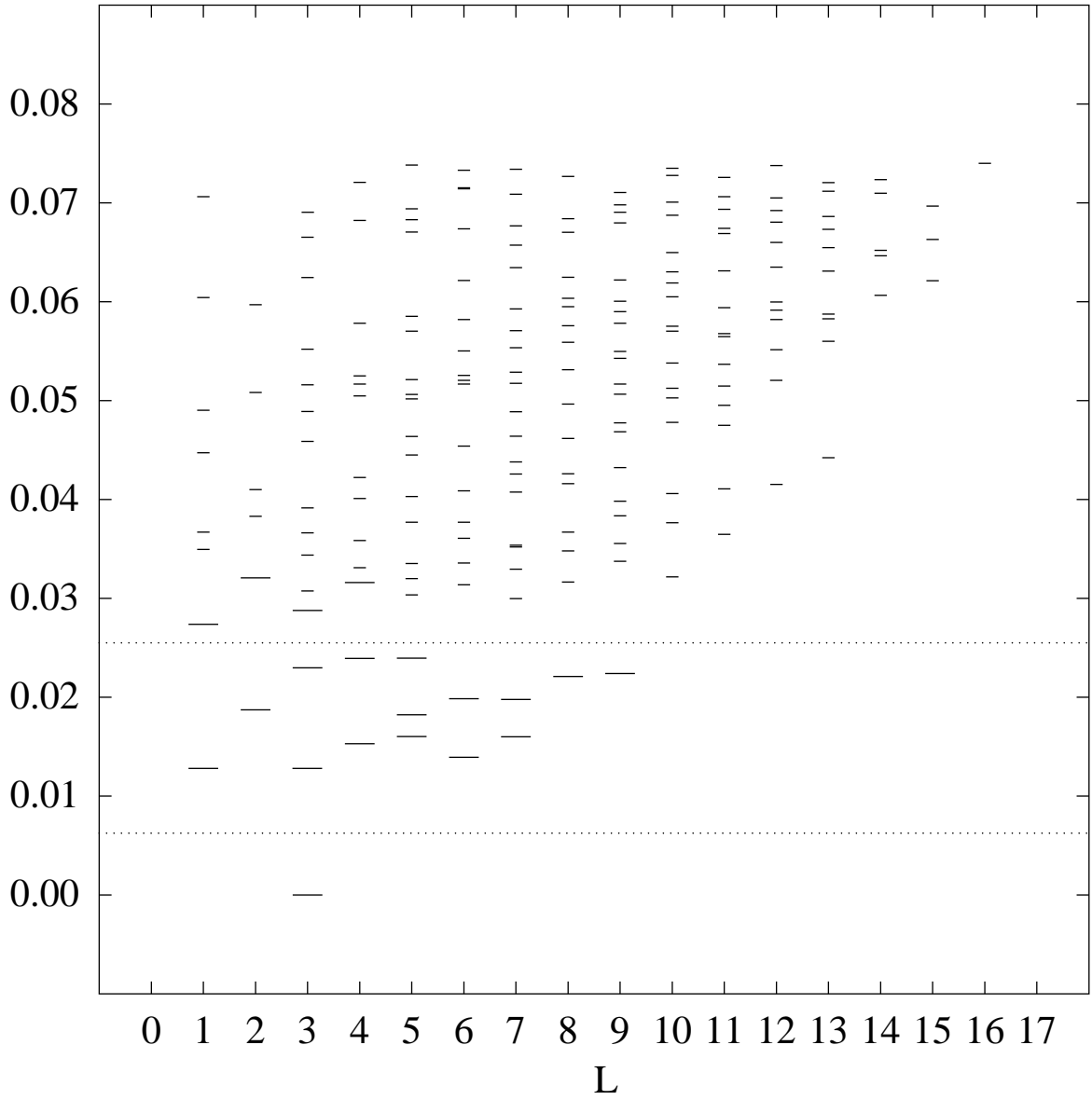
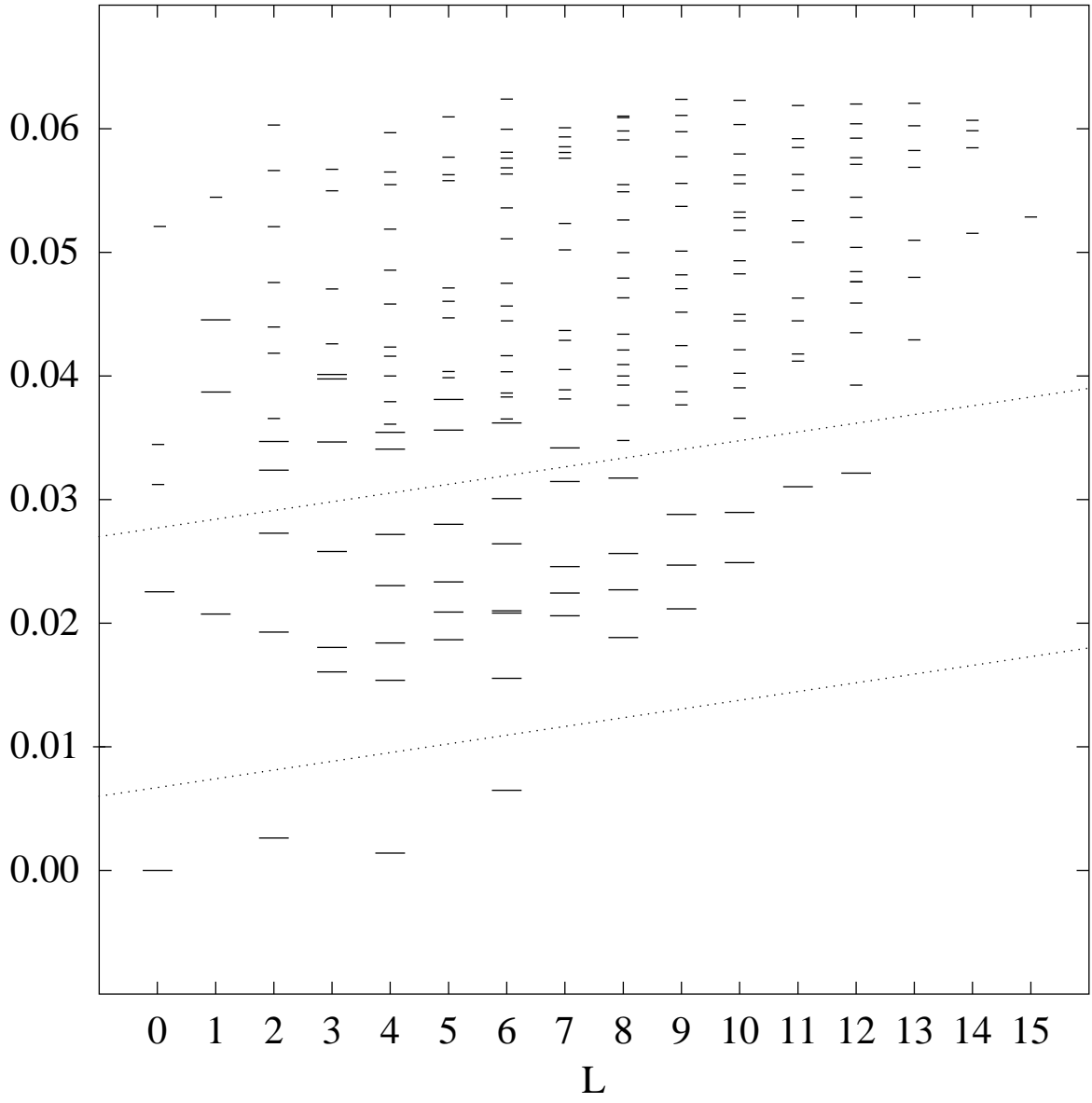
\end{document}